 \DocumentMetadata{
   pdfstandard=A-2b,
   pdfversion=1.7,
   lang=en-US
 }

\documentclass[sigconf, nonacm]{acmart}

\newcommand\vldbavailabilityurl{}
\newcommand\vldbpagestyle{empty} 
\usepackage{xcolor}
\usepackage{balance}

\usepackage{tikz}
\usetikzlibrary{arrows.meta,positioning, calc}

\usepackage{placeins}
\begin{document}
\title{Why We Created Yet Another Memory Framework: Understanding MGA's Role in Next-Gen Database Systems}
\author{Vikramraj Sitpal}
\orcid{0009-0009-4083-9330}
\authornote{These authors contributed equally to this work.}
\affiliation{%
  \institution{Oracle America Inc.}
  \streetaddress{400 Oracle Parkway}
  \city{Redwood Shores}
  \state{California}
  \country{USA}
  \postcode{94065}
}
\email{vikramraj.sitpal@oracle.com}

\author{Pei Li}
\authornotemark[1]
\orcid{0009-0003-3844-8556}
\affiliation{%
  \institution{Oracle America Inc.}
  \streetaddress{400 Oracle Parkway}
  \city{Redwood Shores}
  \state{California}
  \country{USA}
  \postcode{94065}
}
\email{peili.io@oracle.com}

\author{Shubham Kumar}
\authornotemark[1]
\orcid{0000-0001-8674-7962}
\affiliation{%
 \institution{Oracle America Inc.}
  \streetaddress{400 Oracle Parkway}
  \city{Redwood Shores}
  \state{California}
  \country{USA}
  \postcode{94065}
}
\email{shubham.ac.kumar@oracle.com}

\author{Somansh Reddy Satish}
\authornote{Author was at Oracle America Inc. during the time of paper preparation.}
\affiliation{%
  \institution{Oracle America Inc.}
  \streetaddress{400 Oracle Parkway}
  \city{Redwood Shores}
  \state{California}
  \country{USA}
  \postcode{94065}
}
\email{somanshreddy@gmail.com}

\author{Ravi Thammaiah}
\affiliation{%
  \institution{Oracle America Inc.}
  \streetaddress{400 Oracle Parkway}
  \city{Redwood Shores}
  \state{California}
  \country{USA}
  \postcode{94065}
}
\email{ravi.thammaiah@oracle.com}

\author{Nagarajan Muthukrishnan}
\affiliation{%
  \institution{Oracle America Inc.}
  \streetaddress{400 Oracle Parkway}
  \city{Redwood Shores}
  \state{California}
  \country{USA}
  \postcode{94065}
}
\email{nagarajan.muthukrishnan@oracle.com}


\begin{abstract}
Despite the presence of multiple memory regions in modern database systems, supporting an efficient form of memory remains a challenge under production constraints. In enterprise-grade data systems, existing abstractions impose a trade-off between coarse-grained global sharing and strict process isolation, resulting in data copying, memory fragmentation, and limited support for controlled sharing. These challenges become more pronounced as workloads grow more diverse, and systems must tolerate process failures while maintaining predictable performance.

This paper introduces the Managed Global Area (MGA), a scoped shared-memory abstraction in Oracle AI Database that addresses these limitations. MGA allows components to explicitly define allocation source, membership, and coordination semantics across selected processes while integrating with a production database engine. Unlike fully shared memory regions in Oracle, such as the System Global Area (SGA), MGA supports dynamic process membership and modular memory usage without imposing system-wide visibility.

We evaluate MGA on analytical and AI workloads that stress shared-memory execution, including TPC-H hash joins and ONNX Runtime inference. Under concurrent execution, MGA reduces latency for join-intensive TPC-H queries by up to 35\%. For ONNX-based inference, MGA-enabled model sharing reduces memory footprint by up to 90\% and lowers large-model inference latency by up to 37\%. These results demonstrate that dynamically scoped shared memory can improve both efficiency and predictability in production database systems.
\end{abstract}
\maketitle
\pagestyle{\vldbpagestyle}
\begingroup
\renewcommand\thefootnote{}\footnote{
\noindent
\textit{This is the authors’ accepted manuscript. The final version will appear in the Proceedings of the VLDB Endowment (PVLDB), 2026.}\\
This work is licensed under the Creative Commons BY-NC-ND 4.0 International License. Visit \url{https://creativecommons.org/licenses/by-nc-nd/4.0/} to view a copy of this license. For any use beyond those covered by this license, obtain permission by emailing \href{mailto:info@vldb.org}{info@vldb.org}. Copyright is held by the owner/author(s). Publication rights licensed to the VLDB Endowment. \\
}\addtocounter{footnote}{-1}\endgroup

\ifdefempty{\vldbavailabilityurl}{}{
\vspace{.3cm}
\begingroup\small\noindent\raggedright\textbf{PVLDB Artifact Availability:}\\
The source code, data, and/or other artifacts have been made available at \url{\vldbavailabilityurl}.
\endgroup
}

\section{Introduction}

In modern database systems, memory is a critical resource that directly determines performance, scalability, and predictability under production workloads. This challenge becomes particularly pronounced in converged database systems such as Oracle, which natively support diverse data types, workloads, and development paradigms within a single unified engine. As database engines scale to hundreds of CPU cores and support increasingly diverse execution models, including OLTP/OLAP, vectorized execution, and multi-tenant concurrency, traditional memory abstractions often become a limiting factor rather than an enabler \cite{castrillon_declarative_2026, scalecache}.

\begin{figure}
  \centering
  \includegraphics[width=0.90\linewidth]{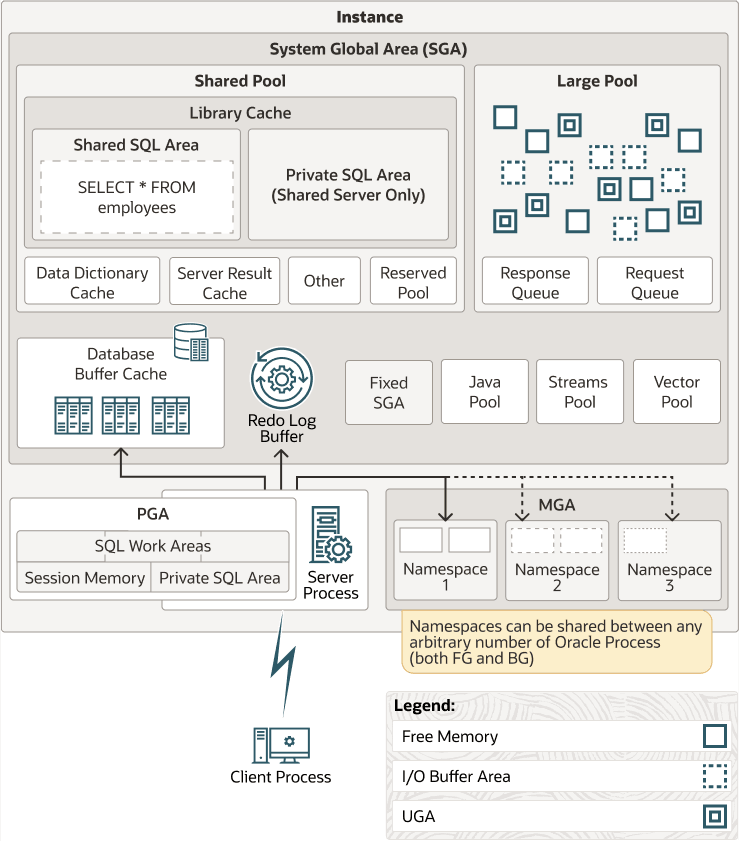}
  \caption{Oracle AI Database Memory Architecture \cite{oracle_db_concepts_2025}}
  \label{fig:instance}
\end{figure}

Oracle AI Database has historically relied on two primary memory areas, the System Global Area (SGA) and the Program Global Area (PGA) \cite{oracle_db_concepts_2025}. As shown in Figure \ref{fig:instance}, the SGA is a fully shared-memory region that is accessible by all server processes, enabling global coordination and inter-process communication. In contrast, the PGA is private to individual server execution entities and is optimized for fast and isolated execution. While this dichotomy has served Oracle well for decades, it introduces fundamental limitations for modern workloads that require selective and dynamic sharing and predictable performance under memory pressure. Beyond simply allocating a shared-memory region, the database must also support how that region is backed, enabling flexible placement across heterogeneous memory types and hierarchies \cite{huang_hash_2026}.

In production environments, these limitations manifest as concrete and recurring challenges. Database components that require controlled sharing among a subset of processes cannot efficiently leverage either SGA or PGA semantics. Using the SGA enforces coarse-grained sharing and static structures, which increase contention and complicate lifecycle management \cite{leis_art_2016}. Relying on the PGA prevents direct sharing and leads to excessive data copying, indirect coordination mechanisms, and fragmented memory usage \cite{dursun_revisiting_2017, oukid_memory_2017}. These inefficiencies become increasingly pronounced as core counts grow and workloads become more skewed \cite{balkesen_multi-core_2013}, ultimately surfacing as tail-latency regressions and reduced throughput under strict service-level objectives \cite{lersch_enabling_2020}.

This is further amplified in converged databases \cite{convergeddbblog}, where heterogeneous components with different lifetimes and isolation needs rely on multiple execution entities that require selective sharing of state without resorting to SGA exposure or costly PGA-based designs with fixed memory-backing semantics.

Existing academic and open-source memory management techniques \cite{ivanova_architecture_nodate, dursun_revisiting_2017, stonebraker1991postgres, raasveldt_duckdb_2019} focus on redesigning engine internals or introducing specialized memory abstractions. Other work advocates tighter database–operating system (OS) co-design to overcome limitations of generic OS abstractions \cite{giceva2011cod, giceva_towards_nodate, kernbypass}. However, in large-scale production systems such as Oracle AI Database, requiring a DB engine redesign or kernel-level modifications is often impractical. Memory abstractions must integrate with a mature and highly optimized codebase, preserve backward compatibility, and operate entirely within user space while maintaining predictable latency behavior \cite{neumann_efficiently_2011, massachusetts_institute_of_technology_end_2018, huang_art_2023}.

Modern database systems face increasing query complexity and growing hardware heterogeneity, which expose limitations in conventional memory management models \cite{huang_hash_2026}. To address these challenges, we introduce the Managed Global Area (MGA), a new memory abstraction in Oracle AI Database.

MGA serves both as an execution-level memory framework and as a scoped shared-memory area. As a framework, it provides a low-level interface that allows database components to explicitly define source, allocation, sharing, and coordination policies across a selected set of trusted processes. As a memory area, MGA represents an instance-scoped shared area that physically stores shared data structures. Its scope can range from a single process to all processes within the database instance. Processes dynamically attach to and detach from MGA, enabling modular, elastic, and on-demand memory usage without imposing global semantics.


Unlike traditional memory systems, MGA provides fine-grained control over memory allocation and process attachment. Memory regions can be selectively attached to specific processes rather than being globally visible. This selective attachment model reduces unnecessary contention on shared structures and limits unintended access, thereby mitigating common risks such as accidental data corruption in shared-memory environments.

One of MGA’s key advantages is its ability to provide contiguous memory allocations, which improves both performance and efficiency by minimizing the overhead associated with fragmented memory. This feature is particularly important for performance-critical operations, such as hash joins and in-memory analytics, where contiguous memory is essential for fast data processing and minimal latency. Furthermore, MGA offers reduced data copying compared to traditional PGA-based designs, which often require costly memory transfers during query execution. By minimizing the need for data copying, MGA reduces CPU cycles and memory bandwidth consumption, which directly contributes to faster query response times.

In addition, MGA provides enhanced yet configurable memory accounting and improved observability, which are vital for monitoring and optimizing memory usage in a database system. These capabilities allow administrators to gain better insights into memory consumption patterns and fine-tune resource allocation to meet the specific needs of different workloads. MGA also mitigates the common problem of memory fragmentation, a persistent issue in many traditional memory management systems that can degrade performance over time. By offering more efficient memory reuse and recycling, MGA ensures that memory resources are utilized optimally, even in long-running, heavy workloads.

Beyond its immediate benefits for conventional memory architectures, MGA’s framework-based design positions it as a future-proof solution that can extend to emerging memory technologies, such as persistent and other non-volatile memories. Although MGA is primarily designed for local DRAM (and PMEM in the past) in Oracle AI Database, its underlying design principles extend naturally to other memory backends, including remote disaggregated memory and Compute Express Link (CXL)–based memory, as explored in recent work on distributed shared-memory database architectures \cite{huang_hash_2026, wang_case_2022}. Together, these capabilities make MGA not only a critical component for modern database systems but also a foundational building block for future database architectures that must adapt to new hardware advancements and operate efficiently across increasingly heterogeneous memory systems. As memory technologies evolve, MGA’s ability to integrate and manage diverse memory types will help it continue meeting the performance, scalability, and reliability demands of next-generation database applications.

We evaluate MGA using AI inference and analytical workloads, with shared-memory execution identified as the dominant bottleneck. MGA substantially improves latency across both workload classes, delivering up to 35\% speedup on TPC-H join queries and up to 37\% speedup on ONNX-based inference.

This paper makes the following contributions:

\begin{itemize}
\item We identify and characterize fundamental limitations of existing memory areas in Oracle AI Database when supporting modern, large-scale analytical and multi-process workloads under production constraints.
\item We describe the design and implementation of MGA, focusing on the key engineering decisions and trade-offs required to integrate a new memory abstraction into a mature, production database.
\item We discuss experience and lessons learned from using MGA within Oracle AI Database, highlighting implications for future database memory management designs.
\item We evaluate MGA using representative analytical workloads and AI workloads to demonstrate its effectiveness without compromising stability or observability.
\end{itemize}

The remainder of the paper is organized as follows. Section \ref{background} provides background on existing memory areas in modern Oracle AI Database systems and introduces the design rationale of MGA. Section \ref{overview} presents the motivation and an overview of the system architecture, and describes the lifecycle of MGA. Section \ref{implementation} details the implementation of MGA, including memory layout, synchronization, and recovery mechanisms. Section \ref{usecases} presents use cases and experimental evaluation of MGA under realistic operational conditions, including MGA-backed hash join execution on TPC-H, as well as the runtime
and memory efficiency of ONNX model sharing. Section \ref{related-work} discusses related work, and Section \ref{conclusion} concludes with directions for future work.

\section{BACKGROUND} \label{background}
Relational database management systems are highly concurrent platforms that share physical resources with the operating system (OS). The OS allocates and multiplexes cores, memory, and I/O to serve diverse workloads, whereas the database must deliver predictable latency, high throughput, and strong isolation for a specific execution model. As a result, the database often treats the kernel as a provider of basic mechanisms such as scheduling, virtual memory, and I/O primitives, while retaining responsibility for policy decisions that require workload awareness \cite{kernbypass}. Memory management is a central aspect of this tension where generic OS behavior can introduce variability through paging, cache interactions, and allocation granularity. The remainder of this section therefore examines how Oracle AI Database structures and governs memory to reconcile OS mediation with database-level performance and correctness requirements.

Memory in Oracle is organized into logical areas that define how allocations are shared and how long they remain valid. Each area is characterized by its \emph{scope}, which determines visibility across Oracle execution entities, and its \emph{duration}, which determines its lifetime in database terms. An Oracle execution entity may correspond to an OS process, an OS thread, or a user-level thread, depending on the deployment model. Choosing appropriate scope and duration is fundamental to achieving efficient reuse and deterministic reclamation, while preserving isolation and predictability under concurrency.

The memory hierarchy in the Oracle AI Database including major memory areas with their scope and duration is described in \\ Table \ref{tab:memory_areas}.

\begin{table}[h!]
    \centering
    \caption{Characteristics of Memory Areas in Oracle AI Database}
    \begin{tabular}{|c|p{3cm}|p{3cm}|}
        \hline
        \textbf{Name} & \textbf{Scope/Visibility} & \textbf{Lifetime} \\
        \hline
        SGA & Shared & Lifetime of an instance \\
        \hline
        PGA & Private & Lifetime of Oracle Process \\
        \hline
        DGA & \begin{tabular}[c]{@{}l@{}}Shared Across OS \\ threads\end{tabular} 
            & Lifetime of OS Process \\
        \hline
        UGA & Shown in Table \ref{tab:pga-sga-comparison}
            & Lifetime of a user session \\
        \hline
        CGA & Private & Lifetime of a database call \\
        \hline
        MGA & Client defined & Client defined \\
        \hline
    \end{tabular}
    \label{tab:memory_areas}
\end{table}

\subsection{Memory Allocation and Management}
 Oracle AI Database uses an internal heap allocator and metadata management layer that standardizes how the database acquires, organizes, and reclaims dynamic memory across process‑private and shared domains. It exposes a consistent model for building heaps, subheaps, carving allocations, growing capacity under pressure, and instrumenting usage, so higher‑level components can reason about memory with predictable performance and governance. 

A heap is a logical arena made of extents, and extents in turn are carved into chunks. An extent is a contiguous region of address space reserved for the heap and aligned according to policy, coming from a backing memory that could be the System Global Area (SGA), Program Global Area (PGA), Dynamic Global Area (DGA) or any other memory type that provides the address space. A subheap is a heap that is part of a larger heap with its own extents and free-list. This helps reduce fragmentation. For performance, the heap manager employs subpools and per‑latch structures to scale across CPUs and to place extents near the requesting NUMA node when possible.

\subsection{System Global Area}
The System Global Area (SGA) is the shared memory of the Oracle AI Database, mapped into all server processes to host common state and high‑value caches. It provides a single, coherent address space for coordination, allowing sessions and background processes to see and update shared structures without inter‑process copying.

The SGA is initialized with explicit layout and versioning metadata, and each process maintains mapping state to attach and detach safely. Concurrency is mediated through component‑specific latches and recovery policies so shared structures remain consistent under load and faults.

Core components of the SGA include the buffer cache for database blocks; the shared pool, which contains the dictionary cache for frequently used metadata needed during SQL parsing and supports shared allocations; and specialized pools such as the large pool, Java pool, and Streams pool. The dictionary cache, or row cache, stores frequently used metadata for fast SQL parsing. These components are implemented on top of the heap manager, which supplies the allocator semantics, namely, extents, free lists, subheaps, and accounting, while the SGA provides the shared, NUMA‑aware address space and protection.
  
Multitenancy adds per‑tenant isolation and governance within the shared region. In a container database (CDB) with pluggable databases (PDBs), the SGA maintains per‑PDB heaps and pools alongside instance‑wide components, attributes allocations and statistics to the correct PDB, and enforces limits such as the aggregate of PDB‑level minimums. Lifecycle notifiers create and retire each PDB’s SGA structures at open and close, while resource management consults per‑PDB statistics for decisions under pressure. NUMA placement and latching are likewise scoped to reduce cross‑tenant interference.

\subsection{Program Global Area}
The Program Global Area (PGA) is the private memory of a server process and provides a private area for computation and control during database execution. It houses the process stack and runtime control blocks; in dedicated server configurations, each user process is served by one server process. It also carries session state, while large transient structures such as sort and hash work areas are carved from it and spill to temporary storage when they outgrow in‑memory limits. Its scope is confined to a single OS thread, its lifetime matches that thread, and its contents are never visible to other processes. Allocation follows the kernel heap framework so components obtain variable‑sized chunks from per‑process heaps with predictable performance, bulk teardown at process exit, and rich attribution for diagnostics and resource governance. Instance memory management coordinates the balance between private work in the PGA and shared caches in the SGA, tightening or relaxing private work area policies under pressure so the system preserves stability while sustaining throughput.

\begin{table}[ht]
\centering
\caption{Comparison of PGA and SGA vis-à-vis UGA}
\begin{tabular}{|p{3.2 cm}|c|c|}
\hline
 \textbf{Memory Area} & \textbf{Dedicated Mode} & \textbf{Shared Mode} \\
\hline
Location of the fixed area & PGA & SGA \\
\hline
Location of the run-time area for DML and DDL & PGA & PGA \\
\hline
Scope of session memory & Private & Shared \\
\hline
\end{tabular}
\label{tab:pga-sga-comparison}
\end{table}
\subsection{Dynamic Global Area}
The Dynamic Global Area (DGA) is a process‑global, thread‑shared memory region that supports components requiring fast, in‑process sharing without the visibility and coordination costs of instance‑wide shared memory. It is mapped once per operating‑system process, accessible to all threads in that process, and its lifetime is bound to the process rather than to individual sessions or calls. Within this region the database initializes allocator‑backed heaps for structures that must be shared across threads, such as caches, coordination metadata, or transport buffers while avoiding inter‑process synchronization and attachment overhead.

\subsection{User Global Area}

The User Global Area (UGA) is the session‑scoped memory area that persists across calls and encapsulates a session’s runtime state. It's a logical area that holds control blocks and metadata that must survive parse–execute–fetch boundaries, including session and transaction state, cursor and cursor‑cache structures, the package state for PL/SQL, security and National Language Support (NLS) contexts, and references to session‑duration resources such as temporary objects and Large Object (LOB) locators. The UGA’s lifetime is the session’s lifetime: it is created at logon, grows and shrinks as the session allocates and frees state, and is reclaimed deterministically at logoff. Table~\ref{tab:pga-sga-comparison} summarizes the memory area backing the UGA under different server process modes: dedicated (one process per session) and shared (multiple sessions multiplexed onto a single server process).

\subsection{Multi-tenancy from the Lens of Memory Areas}
Oracle’s multitenant architecture partitions the SGA so that each pluggable database (PDB) receives a trackable slice. When a PDB opens, the root container reserves the required capacity, sets up container-specific fixed-memory descriptors, and records the allocation in its bookkeeping tables. 

PGA follows the same dual-ledger philosophy. Each allocation updates both the instance total and the owning PDB’s total, so regulators can address local pressure before escalating to global intervention. Under load, remediation first targets sessions whose tenant has exceeded its allowance, preventing any single database from saturating the host. Background activity can be charged either to the global or tenant totals according to policy, but foreground session work is always attributed to the PDB that requested it. The result is a memory hierarchy that delivers tenant isolation alongside centralized control; essential ingredients for running many independent databases inside one Oracle instance.

\subsection{Workloads with Sophisticated Memory Requirements}

Modern database workloads have shifted from predominantly relational query processing to a heterogeneous mix of analytics, machine learning, high-throughput event processing, and low-latency communication. This evolution has increased both the scale and diversity of memory usage patterns and has exposed limitations in traditional Oracle memory areas such as the SGA, PGA, and DGA. These trends motivate a new memory region designed for these workloads with the following properties:

These workloads are characterized by large working sets, bursty and phase-dependent allocation patterns, concurrent access across multiple execution contexts, strong sensitivity to hardware locality, particularly on NUMA systems, and heterogeneous lifetimes ranging from per-operator to multi-stage or multi-request reuse. They also tend to appear in environments where multiple tenants or services execute concurrently, making attribution and isolation operationally important.

\subsection{Oracle Memory Area Guarantees}
The
Oracle AI Database memory areas such as the SGA, PGA, DGA and UGA are required to satisfy stringent recoverability and availability standards. They maintain recovery structures that capture the precise state of in-flight operations, including transactional progress and relevant process and latch state, so that execution can be reconstructed after errors, process termination, or instance failure. Upon abnormal termination, recovery components consult these structures to drive cleanup and ensure that transactions are completed or rolled back correctly. These mechanisms are essential to preserving correctness guarantees, maintaining data integrity and consistency, and supporting enterprise service-level objectives through rapid and reliable fault recovery.

\section{OVERVIEW} \label{overview}

In this section, we present Managed Global Area (MGA). MGA is a new memory area and a framework in Oracle AI Database. As a framework, it offers a common, low-level interface that lets database modules decide how/where memory is allocated, shared, and coordinated across a selected set of trusted processes. As a memory area, MGA is a scoped shared region within an Oracle AI Database instance that physically stores shared data structures and metadata. Processes dynamically attach to and detach from MGA, enabling configurable, elastic, modular, on-demand, and reusable memory usage.

\subsection{Motivation}

Existing memory areas in Oracle AI Database are optimized for well-defined and long-established usage patterns, but they do not adequately support selective, dynamic sharing among a subset of processes. The SGA provides instance-wide shared memory, enabling global coordination but enforcing coarse-grained visibility and static lifetimes that increase contention and complicate lifecycle management for components that require scoped sharing. In contrast, the PGA is private to a single process, preventing direct sharing and forcing data copying or indirect coordination when multiple processes must operate on the same working set.

Other memory areas address orthogonal needs but remain insufficient for this use case. The DGA enables sharing across threads within a single process but does not extend across process boundaries in multi-process deployments. Session and call scoped areas such as the UGA and CGA are tied to Oracle execution contexts whose lifetimes and ownership models do not align with operator-scoped or component-defined shared
memory. As a result, database components that require large, contiguous memory regions, explicit attach and detach semantics, and predictable behavior under process failure are forced into inefficient or fragile designs when relying on existing memory areas.

These limitations motivate a new abstraction that supports controlled sharing without imposing any constraints whatsoever while remaining compatible with production requirements of robustness and scalability.


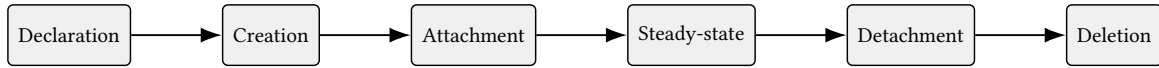
\begin{figure*}[t]
\centering
\begin{tikzpicture}[
  font=\small,
  state/.style={
    draw, line width=0.55pt, fill=black!6,
    rounded corners=2pt,
    minimum height=8mm,
    inner xsep=4pt, inner ysep=2pt,
    align=center
  },
  arr/.style={-{Latex[length=2.8mm,width=2mm]}, line width=0.7pt},
  node distance=12mm
]
\node[state] (decl)  {Declaration};
\node[state, right=of decl]  (create) {Creation};
\node[state, right=of create] (attach) {Attachment};
\node[state, right=of attach] (steady) {Steady-state};
\node[state, right=of steady] (detach) {Detachment};
\node[state, right=of detach] (delete) {Deletion};

\draw[arr] (decl) -- (create);
\draw[arr] (create) -- (attach);
\draw[arr] (attach) -- (steady);
\draw[arr] (steady) -- (detach);
\draw[arr] (detach) -- (delete);
\end{tikzpicture}
\caption{Namespace lifecycle states.}
\label{fig:ns-states}
\end{figure*}
\subsection{Use Cases}
\subsubsection{{Parallel Hash Joins}}
Parallel hash join allocates shared operator state in MGA so that all participating parallel server processes can attach to a single, in-place image of the build and probe structures. At operator open, a named MGA region is provisioned for the query and operator. Parallel builders and probers attach to this region to construct and access hash partitions, bucket arrays, overflow chains, and coordination metadata directly. Builders materialize the hash table once in MGA, and probers read it directly. When radix partitioning is enabled, fan-out buffers, partition headers, and spill bookkeeping are also placed in the same region. As a result, producers and consumers can share partition state and coordinate phase transitions without transferring data between private memory.

\subsubsection{{Open Neural Network eXchange (ONNX)}}
ONNX specifies an open and framework-agnostic model format with a versioned operator set, which enables reliable interchange across toolchains. ONNX Runtime optimizes and executes computation graphs on CPUs, GPUs, and accelerators, thereby providing portable and low-latency inference.
For in-database inference, MGA is used to store model graphs, initializers, and pre-optimized weights in instance-scoped shared memory. Both persistent artifacts and transient working sets allocate from MGA under feature-scoped quotas, which provides governance, isolation, and deterministic reclamation.

\subsubsection{{Hierarchical Navigable Small World (HNSW) Graph Indexing}}
Across the HNSW stack, MGA provides shared and process-independent memory for building, reloading, and maintaining the graph. During index creation and repopulation, a coordinator provisions instance-scoped vector pools in MGA. Parallel workers attach to these pools and allocate vectors, per-layer structures, adjacency lists, and identifier maps in place. Pool-level quotas and growth controls bound consumption by index or snapshot. Freed resources are returned to the shared pool, which reduces churn and fragmentation. Because MGA persists beyond the lifetime of any single process, partial work can resume after faults, and deterministic teardown can occur at phase boundaries.

\subsubsection{Inter-process Communication (IPC) across Remote Instances}
The IPC RDMA subsystem provides high-throughput, low-latency messaging by registering memory for direct network access using a single large region per NUMA socket. This region is pre-registered with the network interface, allowing server processes to sub-allocate buffers without re-registration.

IPC uses MGA to streamline RDMA by provisioning a single instance-scoped segment, registered once via IPC0 coordinator. All processes attach to this segment and allocate buffers within the shared space. MGA supports large, aligned memory regions to improve DMA and cache efficiency. The segment can use huge pages and prefetch hints to reduce TLB pressure. Governance mechanisms ensure isolation, while centralized registration and cleanup enhance stability, reduce latency, and boost RDMA throughput.

\subsubsection{Session Heap for Shared Server}
In shared-server architecture, each session has a persistent working set that must be accessible to any worker process. Storing this per-session heap (UGA) in MGA materializes it as an instance-scoped, attachable area. Dispatchers and workers attach as needed, execute, and detach. Lifecycle management, accounting, and reclamation are governed centrally at session termination, not by individual workers. This approach avoids relocating or duplicating session state during process switches and decouples memory ownership from transient workers.

Using MGA lowers cache contention and fragmentation by offloading session data, while supporting governance through namespaces, quotas, and tenant-level accounting. MGA provides large, aligned regions with optional huge-page backing and NUMA locality, improving cache and TLB efficiency for sessions with intensive workloads. Centralized allocation and teardown ensure deterministic cleanup, efficient worker churn and reconfiguration, and uniform observability. These benefits reduce latency and improve throughput in shared-server environments.

\subsubsection{{Cursor Duration Temporary Tables}}
Cursor Duration Temporary Tables (CDTs) that store in-memory content utilize MGA as a tenant-scoped workspace, created for the lifespan of a cursor execution and shared by all participating processes. For each execution, a dedicated MGA region is allocated per PDB and cursor, accompanied by lightweight metadata covering ownership, accounting, and coordination. Parallel loaders and scanners are admitted via attach and detach actions. CDT rows are materialized in place. After execution, all participants detach, and the region is reclaimed in one step, which enforces deterministic teardown and eliminates cleanup inconsistencies. Under memory pressure, CDTs automatically spill to temporary space to free MGA capacity while maintaining correctness.

\subsubsection{{OTLP-Based Observability}}
The distributed tracing subsystem, using the OpenTelemetry Protocol (OTLP) \cite{oracle_dbms_observability_26c}, uses MGA to store operation metadata and trace data in instance-scoped shared memory, supporting concurrent producers and archival consumers without duplication. MGA namespaces, which can be sharded for locality, allow processes to dynamically attach and detach while emitting or draining traces. Each namespace manages pooled trace buffers, enabling zero-copy export and efficient background archival. Governance mechanisms based on namespaces, quotas, and accounting enforce predictable memory limits and safe reconfiguration as processes join or leave. The trace bucket service allocates pools from MGA, ensuring bounded usage and controlled overflow behavior, while per-process defaults provide uniform policy and visibility.

\subsection{Design Goals}
\label{subsec:mga-design-goals}

MGA is designed to fill this gap by introducing a scoped shared-memory abstraction with explicit lifecycle and membership semantics. It is guided by the following design goals:

\begin{itemize}
  \item \textbf{Selective sharing.} Define precise attachment and membership semantics so memory is shared only among an intended subset of trusted processes, with explicit grant/revoke behavior.

  \item \textbf{Elastic memory.} Support online growth and shrinkage of shared regions over time, adapting to workload dynamics without restart-based reprovisioning.

  \item \textbf{Decoupled lifetime.} Support shared memory lifetimes independent of individual process lifetimes and not rigidly tied to database instance uptime, enabling controlled reuse and deterministic teardown.

  \item \textbf{Pluggable memory backend.} Separate shared memory management policy from the mechanisms that supply physical memory, allowing multiple backends without duplicating core lifecycle, protection, and accounting logic.

  \item \textbf{Low coordination overhead.} Minimize global contention and coordination on the steady-state fast path so performance scales with increasing process counts.

  \item \textbf{Fault tolerance.} Provide well-defined consistency and recovery semantics under partial failures (e.g., crashes or termination) to prevent leaks, corruption, or permanent inaccessibility.

  \item \textbf{Observability.} Expose statistics and attribution for production monitoring, debugging, and governance (e.g., usage by workload/namespace and lifecycle state).
\end{itemize}

MGA realizes these goals through a set of abstractions that define how shared memory is created, attached to processes, and recovered under failure. The remainder of this section introduces these abstractions including namespaces and segments, and explains how they provide selective sharing with explicit lifecycle and membership semantics.

\subsection{Design}

\subsubsection{Namespaces}
Namespaces are the top-level organizational abstraction in MGA. A namespace defines a
named shared-memory domain within a database instance. Each namespace is a logical container that groups one or more memory regions, known as segments. Together, these form the shared address space exposed to participating processes. A process must explicitly attach to a namespace in order to access its memory. When the process no longer requires access, it detaches from the namespace. 

\subsubsection{Segments}
A segment is a contiguous region of virtual address space carved from a namespace’s reserved address range and is the concrete unit that holds the shared data. The memory backing for a segment in a namespace (e.g., POSIX \texttt{/dev/shm}, persistent memory, etc) is selected according to namespace type. Segments may be added to or removed from a namespace dynamically. All attached processes can discover and synchronize with the modified layout.

\subsubsection{Namespace Lifecycle}
Figure \ref{fig:scheduler} illustrates the lifecycle of an MGA namespace from the perspective of participating processes. A namespace progresses through a sequence of well-defined stages, while multiple processes may attach to it and detach from it over time.

\begin{enumerate}
  \item \textbf{Declaration.}
  A namespace is declared at compile time, prior to instantiation. At declaration, properties like the namespace’s name, memory backend type, management method, synchronization policy and other attributes are registered.

  \item \textbf{Creation.}
  At runtime, one or more instances of a declared namespace may be created. Creation allocates the namespace’s shared state, which stores namespace-wide metadata, and returns an opaque shared handle to the caller. This also makes the namespace discoverable to processes for attachment.

  \item \textbf{Attachment.}
  Processes attach to the namespace to obtain access. Attachment enrolls a process in the namespace,
  constructs its private (process local) state, and returns an opaque private handle.

  \item \textbf{Steady-State Operation.}
  While the namespace is active, attached processes may query namespace attributes and create or delete
  segments. These operations update shared namespace state and allow processes to reconcile their private
  state according to the namespace’s synchronization policy.

  \item \textbf{Detachment.}
  A process may detach from the namespace at any time. Detachment revokes the process’s access, dismantles
  its private state, and releases process-local resources without affecting other attached processes.

  \item \textbf{Deletion.}
  When the namespace is no longer needed, it is deleted. Deletion blocks new attachments, ensures that no
  attached processes remain, reclaims all associated segments, and releases shared metadata. After deletion completes, the namespace instance ceases to exist, while its declaration remains available for future creation.
\end{enumerate}


\begin{figure}[t]
\centering
\begin{minipage}{0.9\linewidth}
\begin{verbatim}
Process A:
  shared_handle = CreateNamespace(namespace_id, 
                                  name, policy)
  private_handle = Attach(shared_handle)

  seg_addr = CreateSegment(private_handle, size)
  ... use segment(s) ...
  DeleteSegment(private_handle, seg_addr)

  Detach(private_handle)
  DeleteNamespace(namespace_id)

Process B:
  private_handle2 = Attach(shared_handle)
  seg_addr2 = QuerySegment(private_handle2, seg_id)
  ... use segment(s) ...
  Detach(private_handle2)
\end{verbatim}
\end{minipage}
\caption{Client Usage}
\label{fig:scheduler}
\end{figure}

\subsubsection{Pluggable Memory Backend}
MGA defines a pluggable memory backend by abstracting the mechanisms used to provide physical memory away from namespace and segment management. Each namespace binds to one memory backend, and all segment operations within that namespace rely on the selected backend.

This abstraction reflects a deliberate separation of concerns. Namespace and segment lifecycles, synchronization, and recovery are implemented once in MGA and do not depend on how memory is allocated, mapped, or released. Memory-specific behavior is confined to a clear boundary. Adding or removing a memory backend therefore does not require changes to these core algorithms or metadata structures.

As a result, MGA functions not only as a shared-memory area but also as a memory framework. It provides a stable, reusable foundation on which multiple namespaces, each potentially backed by a different memory backend, can coexist within the same database instance.

\section{IMPLEMENTATION} \label{implementation}

This section describes how MGA is implemented in a production database system, including reserving virtual address space, synchronizing segments across processes to maintain a consistent view, providing low-overhead access to shared state, and ensuring correctness under failures.

\subsection{Address Space Reservation (ASR)}

MGA uses an address-space reservation facility to guarantee that large, aligned virtual regions are available before any shared segment is mapped. The facility operates on a dedicated interval of virtual memory that is earmarked for MGA use during system start-up. At that time the platform supplies three inputs: (1) the lower and upper bounds of the permissible address interval, (2) the reservation unit size, typically derived from the OS’s supported page sizes, and (3) any per-instance offset needed in clustered deployments so that separate instances draw from disjoint virtual ranges. This interval is represented as a bitmap in which each bit denotes one reservation unit; a zero bit indicates that the corresponding slice of the address space is free, whereas a one bit indicates that it has been claimed.

Reserving address space proceeds as a deterministic five-stage state machine:

\begin{enumerate}
  \item \textbf{Normalize the request.}
  The caller’s requested length is rounded up to an integral number of reservation units so that every allocation respects the predefined granularity.

  \item \textbf{Select a page size.}
  If the caller prefers a specific system page size, the reservation service filters the platform’s page-size table to those that are exact multiples of the reservation unit. When such a page size is available, the forthcoming reservation is aligned accordingly; otherwise, the default unit size is used.

  \item \textbf{Search for capacity.}
  The bitmap is scanned for a contiguous run of zero bits whose length matches the normalized request. The search operates entirely within the MGA interval and therefore cannot collide with unrelated allocations.

  \item \textbf{Commit the reservation.}
  Once a suitable run is found, the corresponding bits are atomically set while holding a short-lived latch that serializes concurrent reservations. The virtual start address is computed from the base of the MGA interval plus the index of the first reserved unit.

  \item \textbf{Return the address.}
  The aligned address is handed back to the caller, which may now instantiate segments with the assurance that the virtual range will remain unoccupied until explicitly released.
\end{enumerate}

Releasing a reservation reverses the process. The caller supplies the start address and length, the service converts them back into unit indices, acquires the same latch, clears the associated bitmap entries, and returns the range to the free pool. Because the bitmap is the sole authority on which regions are in use, the release is immediate and deterministic.

This design offers two main benefits. First, namespaces that require strict alignment such as those targeting persistent memory, direct I/O segments, or large-page heaps can obtain a guaranteed address range before committing physical resources. Second, in clustered configurations every instance applies the identical procedure but operates on a distinct slice of the MGA interval, ensuring that virtual addresses exchanged across nodes never overlap. Address Space Reservation therefore converts what would otherwise be opportunistic address acquisition into a controlled, reproducible step in the MGA lifecycle.

\subsection{Namespaces}

\subsubsection{Shared vs. Private State}
To maintain consistency across processes, MGA separates namespace state into shared and per-process state. The \emph{shared state} is the authoritative record of a namespace. It stores the namespace’s name, type, an array of segments, statistics, and synchronization primitives. All namespace-wide decisions are made against this shared state. Each attached process maintains a \emph{private state} that mirrors relevant portions of the shared state while tracking process-local state. The private state records which segments are currently mapped, the cleanup actions required upon detachment, and so on. This separation allows MGA to update shared state once while allowing each process to update its local state independently.

\subsubsection{Process membership}
Each namespace maintains a \emph{process list} of attached processes within its \emph{shared state}. To avoid global serialization, the shared state stores the process list as an array of bucketed lists rather than a single list. The number of buckets is configurable, with a system-defined default that may be overridden. Each bucket consists of a header anchoring a doubly linked list of process descriptors and is protected by its own latch. When a process attaches to a namespace, MGA allocates a process descriptor and links it into one of the buckets. The shared state thus maintains an authoritative view of all attached processes. By striping membership across multiple independently latch-protected buckets, MGA allows concurrent attach and detach operations to proceed in parallel when they target different buckets.

\subsection{Segments}

\subsubsection{Shared vs.\ Private Segments}

MGA keeps two views of every segment: a shared segment and a private segment. This split keeps a single authoritative state while letting each process carry only the local state it needs to stay consistent.

The shared segment, stored in the namespace’s shared state, is the authoritative representation. It defines the segment’s lifecycle state, address range, backing store, and generation counter. All namespace-wide decisions about segment creation, deletion, and visibility are made against the shared segment.

Each attached process maintains a corresponding private segment in its private state. The private segment array mirrors the shared segment table but records only process-local state, including whether the segment is unmapped/mapped, the process-local mapping address and size, backend-specific attach state, NUMA placement data relevant to the local mapping, and a private generation counter.

\subsubsection{Lifecycle}
Each segment transitions through different states: \emph{free}, \emph{initialized}, \emph{created}, \emph{mapped}, and \emph{delete-in-flux}. A segment begins in the free state, where it exists in the namespace’s tables but has no address or backing assigned. A request for space moves the segment to initialized, where policy choices are recorded and a range within the namespace’s reservation is set aside, but no memory has yet been allocated.

Once backing storage is successfully obtained from the selected provider, the segment enters the created state. Its base address and size are fixed, version counters advance, and the segment becomes visible to attached processes, although no mappings are installed. The segment enters the mapped state when a process establishes a mapping explicitly or when delayed synchronization handles the first access; from that point, the segment becomes part of the process’s working set.

When the namespace reclaims the space, it marks the segment delete-in-flux. This blocks new attachments and forces existing mappings to be torn down, causing the fault handler to perform a clean detach rather than remapping. After all processes have detached, MGA releases the backing storage, clears the segment metadata, and returns the segment to the free state for reuse.

\subsection{Synchronizing segments}

Synchronization ensures that each process’s view of an MGA namespace stays consistent with its actual contents. As segments are created or deleted, processes must stay updated on which regions exist and how they are mapped. Otherwise, conflicts can arise, like a process using memory already reclaimed or differing segment sizes/ranges.

MGA provides three synchronization policies: eager (syncs all processes immediately), delayed (syncs only when a process accesses the segment), and explicit (puts the responsibility on the client). These are set during namespace declaration.
To detect when synchronization is needed, MGA uses a generation counter in both shared and private states. If a process’s local counter differs from the shared counter, it updates its view of the namespace.

Reconciliation uses a two-phase process to ensure safe transitions between segment layouts. First, the local segment list is aligned with the shared list. Then, two sweeps are done: one to detach outdated mappings, and another to attach new ones. This ensures no address is reused prematurely. In eager sync, any change triggers immediate reconciliation for all processes. In delayed sync, updates are published but reconciliation happens only when a process accesses an address which belongs to a \textit{new segment}. Periodic sweeps keep the namespace consistent. Explicit sync relies on the client to manually trigger updates, with no automatic synchronization.

\subsection{Fixed Variables}
Fixed variables are namespace-scoped shared objects whose layout and addresses are fixed when a namespace is created. Each fixed variable declaration specifies its type, size, and namespace, and is assigned a fixed offset within a dedicated segment reserved for fixed variables. The motivation for fixed variables is to provide an anchor variable.
Fixed variable offsets remain constant for the lifetime of the namespace. When a namespace is created, MGA allocates the fixed-variable segment and records its base address in the namespace’s shared state. Every process that attaches to the namespace maps this segment at the same virtual address. As a result, the address of each fixed variable is identical across all participants and can be computed directly by adding the compile-time offset to the segment base, without coordination. The segment that holds the fixed variables participates in the namespace lifecycle and recovery in the same manner as ordinary segments. During namespace deletion, however, this fixed-variable segment is unmapped only after all other segments have quiesced. This ordering prevents processes from retaining stale pointers into the fixed-variable region while dynamic segments are being torn down.

This design generalizes the benefits of fixed SGA. Like fixed SGA, fixed variables provide stable addresses and safe pointer storage. Unlike fixed SGA, however, fixed variables are scoped to individual namespaces rather than being global to the database instance.

\subsection{Access and Allocation Methods}

Processes interact with MGA exclusively through opaque handles. The shared handle is the token that identifies a namespace instance and can be handed across process boundaries. Attaching converts it into a private handle, which exposes the namespace’s operations while remaining scoped to the process that owns it. Allocation is layered on top of this hierarchy. Each namespace ships with its own managed heap, if configured, so client code calls into MGA to request memory rather than mapping raw segments. MGA sets up subheaps logically so memory lifetime matches the owning workload. Allocations draw from the namespace’s elastic segments; freeing returns space to the same heap, and MGA updates its accounting tables to track per-process and per-group usage. Because every access flows through the handle and the managed heap, MGA can enforce limits, apply checks, and integrate with higher-level resource management without exposing unmediated memory.

\subsection{Recovery}

MGA is designed to tolerate interruptions during all state-mutating operations, including namespace creation and deletion, process attachment and detachment, private-state management, and address-space reservation. To achieve this, MGA associates every latch-protected structure with a \emph{recovery record} that captures the intent and intermediate state of an operation before any irreversible change is performed.

A recovery record encodes the operation type, relevant pointers, and any partially prepared objects required
to complete or undo the operation. If an execution path must temporarily release locks, encounters an error, or is interrupted by process termination, the corresponding recovery routine replays the recorded action to completion or safely rolls it back.

\textbf{Shared-State Recovery.}
Each namespace instance maintains recovery records within its shared state to protect namespace-wide operations, including creation, deletion, and segment updates. These records capture the intended operation and any prepared state before the shared state is linked into or removed from global lists. If allocation or initialization must occur outside the shared latch, the recovery record preserves progress so that cleanup or retry can safely complete once the latch is reacquired.

\textbf{Process-List Recovery.}
Each process-list bucket carries its own recovery record to guard membership changes. During attachment, the
record stages the addition of a process descriptor before the descriptor is linked into the list; during
detachment, it stages unlinking and disposal as separate steps. Because the record resides with the bucket,
cleanup handlers invoked during process failure or system shutdown can complete or undo in-flight membership
changes without corrupting the list.

\textbf{Private-State Recovery.}
Per-process private state maintains its own recovery records to coordinate allocation, linkage, and deletion across threads. Attachment populates the record before the private state becomes visible; detachment reuses the record to orchestrate removal. Any interruption such as explicit abort, thread exit, or background cleanup invokes the same recovery routine to converge the private and shared views to a consistent state.

\textbf{Normal Detach vs. Forced Cleanup.}
During a normal detach, recovery records ensure ordered progress: private-state cleanup precedes removal from the process list, and completion is signaled only after both records return to a no-op state. If detach is interrupted, subsequent entry points invoke recovery routines to finish the staged steps. In forced cleanup scenarios such as abrupt process termination or instance shutdown, MGA inspects recovery records embedded in shared state and process-list buckets. Any record indicating an incomplete operation is replayed, completing pending attachments, undoing partial removals, and freeing allocated resources. As a result, MGA guarantees that namespaces converge to a clean and consistent state even after unexpected failures.

By recording intent rather than outcomes, MGA ensures that all namespace mutations are recoverable, allowing both routine operation and exceptional failure paths to preserve consistency.

\subsection{Statistics and Observability}

MGA provides a multi-layered observability framework designed to support diagnosis, capacity planning, and performance analysis of memory usage. Statistics are exposed at namespace, segment, and operation granularity, allowing clients to reason about both steady-state behavior and dynamic events. Statistics are updated on certain events and refreshed periodically to ensure consistency. Diagnostic interfaces allow clients to snapshot and expose the current state and summary.

\textbf{Namespace-Level Statistics.}
Each namespace maintains statistics such as the number of active segments, the total bytes reserved, page sizes, and memory usage. The namespace-level view enables clients to identify memory-heavy namespaces or track growth over time. To ensure correctness after recovery or delayed cleanup, namespace statistics are periodically recomputed from the namespace metadata.

\textbf{Global Aggregation.}
MGA aggregates per-namespace statistics along multiple dimensions, such as namespace type. These aggregates provide instance-wide summaries that help capacity management without requiring inspection of individual namespaces.

\textbf{Operation-Level Statistics.}
To capture dynamic behavior, MGA instruments segment lifecycle operations such as creation, deletion, attachment, and detachment. For each operation class, MGA records cumulative counts, total elapsed time, and total data volume processed. These counters allow derivation of average latency and payload size, and expose performance anomalies such as spikes in synchronization or allocation overhead.

Together, these mechanisms ensure that MGA remains observable under production workloads, enabling the database to detect memory pressure, diagnose synchronization costs, and understand allocation behavior without intrusive instrumentation.

\section{EXPERIMENTAL EVALUATION} \label{usecases}

This section first outlines representative MGA use cases within Oracle AI Database. MGA supports these use cases through a combination of scalable sharing and efficient memory layout: cached attachment state and namespace sharding reduce coordination overhead and contention, while large, alignment-friendly allocations improve locality and CPU cache behavior. Together, these capabilities enable low-latency data ingestion, zero-copy export across execution contexts, and predictable, instance-level control over memory consumption.

It then presents an experimental evaluation of a selected subset. The results show that MGA improves execution efficiency under concurrency while maintaining predictable performance and efficient resource management.

\subsection{Evaluation Overview}  \label{evaluation}
We evaluate MGA to demonstrate its effectiveness in accelerating hash join execution in analytical workloads. All experiments were conducted on an Oracle Exadata~X8 server. 

For analytical workloads, our evaluation focuses on the TPC-H benchmark \cite{tpch}, considering the subset of queries that are eligible for hash joins. Among the 22 TPC-H queries, 12 queries can be executed using hash joins. These queries form the basis of our end-to-end analytical evaluation. We compare MGA-backed hash join execution against a baseline
that allocates hash join state in PGA, which is private to
each worker process. Under this baseline, hash join execution incurs redundant memory
allocation and increased data movement across worker processes. By contrast, MGA
enables hash join operators to place shared data structures in instance-scoped shared
memory, reducing memory duplication and data exchange overhead. We do not compare MGA
against non-Oracle baselines, as such approaches violate the production constraints
discussed in previous sections.

In addition to analytical workloads, we evaluate MGA using AI inference workloads that
stress shared-memory execution. We consider four representative embedding models:
\emph{all-MiniLM-L12-v2} (a lightweight monolingual text embedding model),
\emph{clip-vit-base-patch32} (a multimodal vision--language embedding model),
\emph{multilingual-e5-base} (a multilingual text embedding model), and
\emph{e5\_large\_v2} (a large-scale, high-quality text embedding model). The workload
invokes ONNX-based embedding inference during query execution, with input text lengths
ranging from 100 to 1000 characters.


In the baseline configuration, each database server process independently loads the ONNX model and allocates private runtime state, resulting in redundant model instances and repeated initialization overhead as concurrency increases. In the MGA configuration, ONNX model graphs and pre-optimized weights are loaded once into MGA memory and reused across processes, enabling zero-copy access. Transient runtime buffers are allocated from MGA, and inference requests are multiplexed through ONNX Runtime.

\textbf{Basic Configuration.}
Experiments were conducted on a single Exadata X8 server with the following configuration:
\begin{itemize}
    \item \textbf{CPU:} 2 x 24-core Intel Xeon Platinum 8260 Processors @ 2.4 GHz;
    \item \textbf{RAM:} 384 GB RAM (12 x 32 GB DIMMs);
    \item \textbf{OS:} Oracle Linux 7 (Unbreakable Enterprise Kernel 4) with Linux kernel 4.14.
\end{itemize}
\subsection{End-to-End Performance on TPC-H}

\begin{figure}
  \centering
  \includegraphics[width=0.7\linewidth]{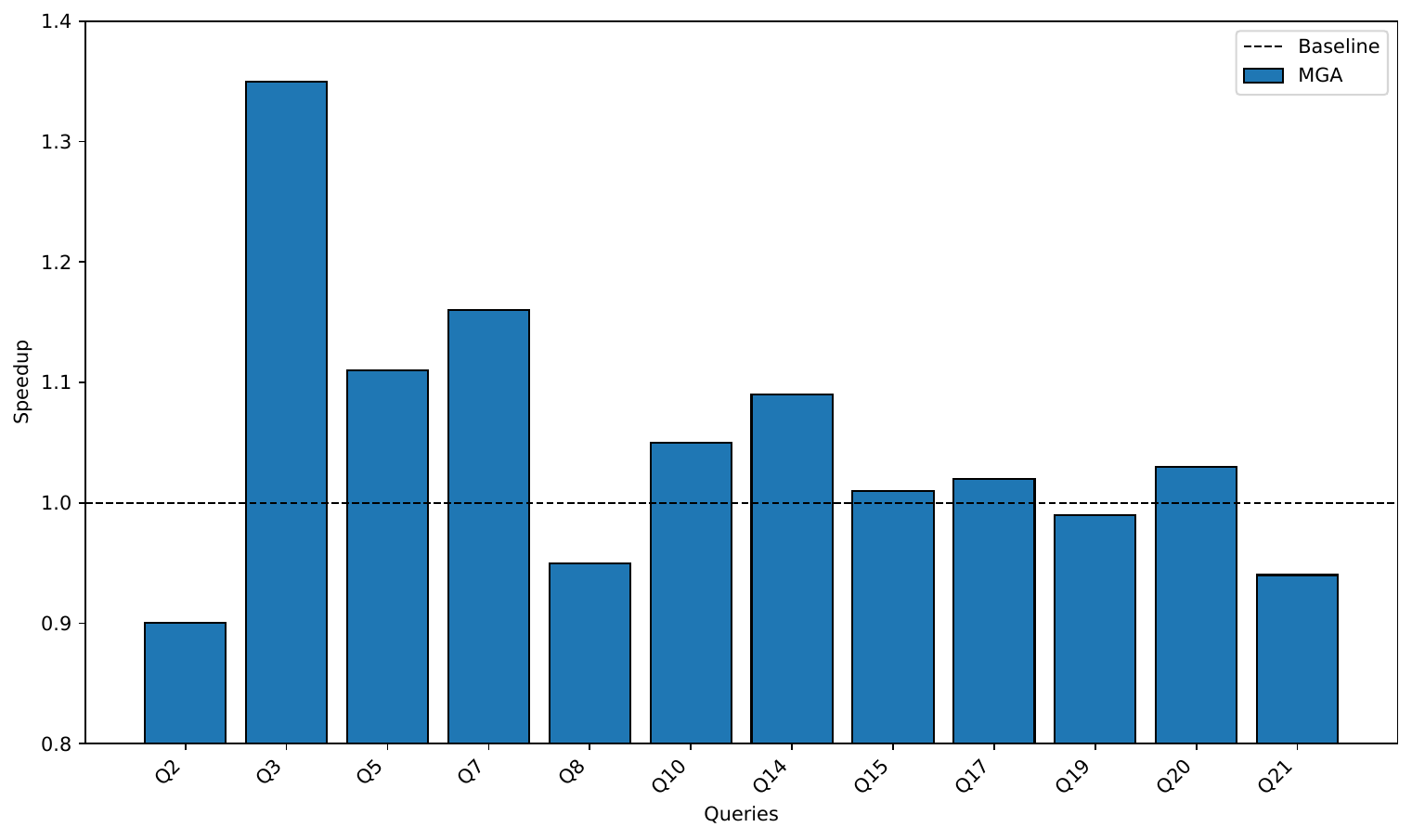}
  \caption{TPC-H performance comparison with and without MGA-backed hash join.}
  \label{fig:tpch}
\end{figure}

In this section, we evaluate the performance improvements enabled by MGA-backed hash join execution on TPC-H workloads. Figure~\ref{fig:tpch} reports the end-to-end query performance under the baseline and MGA configurations.

As shown in \autoref{fig:tpch}, out of the 12 queries that are eligible for hash joins and make use of the MGA use case, performance improvements were observed for the following queries: Q3 (35\%), Q5 (11\%), Q7 (16\%), Q10 (5\%), Q14 (9\%), Q15 (1\%), Q17 (2\%), and Q20 (3\%). Conversely, performance degradation was noted for queries Q2 (10\%), Q8 (5\%), Q19 (1\%), and Q21 (6\%). Further analysis identified that the performance losses were attributed to suboptimal query plan choices made by the query optimizer.

From the results, we make the following observation.
Under concurrent query execution, the baseline configuration incurs increasing overhead due to two sources: (1) repeated memory allocation for per-process hash tables and intermediate buffers, and (2) repeated data copying across processes during intermediate result exchange. As the number of concurrent queries increases, these redundant memory operations amplify memory pressure, degrade cache locality, and introduce significant inter-process communication overhead.

MGA addresses these inefficiencies by enabling hash join operators to allocate and access shared data in instance-scoped shared memory. Moreover, intermediate buffers can be shared across server processes without additional copying or serialization, substantially reducing inter-process data movement. These benefits become more evident at higher concurrency levels: reduced allocation overhead lowers memory pressure, while direct shared-memory access improves cache locality and parallel execution efficiency. Consequently, MGA achieves higher throughput and better scalability under high-concurrency workloads.

Overall, these results show that MGA-backed hash join execution improves TPC-H query performance by addressing two fundamental bottlenecks in concurrent execution: redundant memory copying and excessive data movement. By providing a shared-memory substrate with predictable allocation and lifecycle management, MGA enables hash join operators to scale efficiently under high concurrency without introducing additional coordination overhead.

Despite overall improvements, some queries still suffer from performance degradation. The following summarizes the causes and potential optimizations.

For Queries 2, 8, and 21, we observe performance degradations associated with certain optimizer decisions that interact unfavorably with query characteristics. In Queries 2 and 8, the selection of buffering hash joins correlates with higher memory and CPU usage compared to alternative join strategies under the observed workload. In Query 21, the behavior of \texttt{GROUP BY} pushdown optimization appears to reduce effective parallelism and increase intermediate processing overhead. These cases suggest that, while generally beneficial, specific optimization rules may introduce additional costs under particular data distributions and execution conditions.

\subsection{Runtime Overhead of ONNX Integration} \label{runtime-overhead-of-onnx-model}
This section and the next evaluate the performance of ONNX model sharing via MGA. Here, we measure end-to-end execution time and report runtime overhead relative to the baseline configuration, while the next section focuses on memory usage. We use the same workload described in Section \ref{evaluation} and keep all parameters fixed.

\begin{figure}
  \centering
  \includegraphics[width=\linewidth]{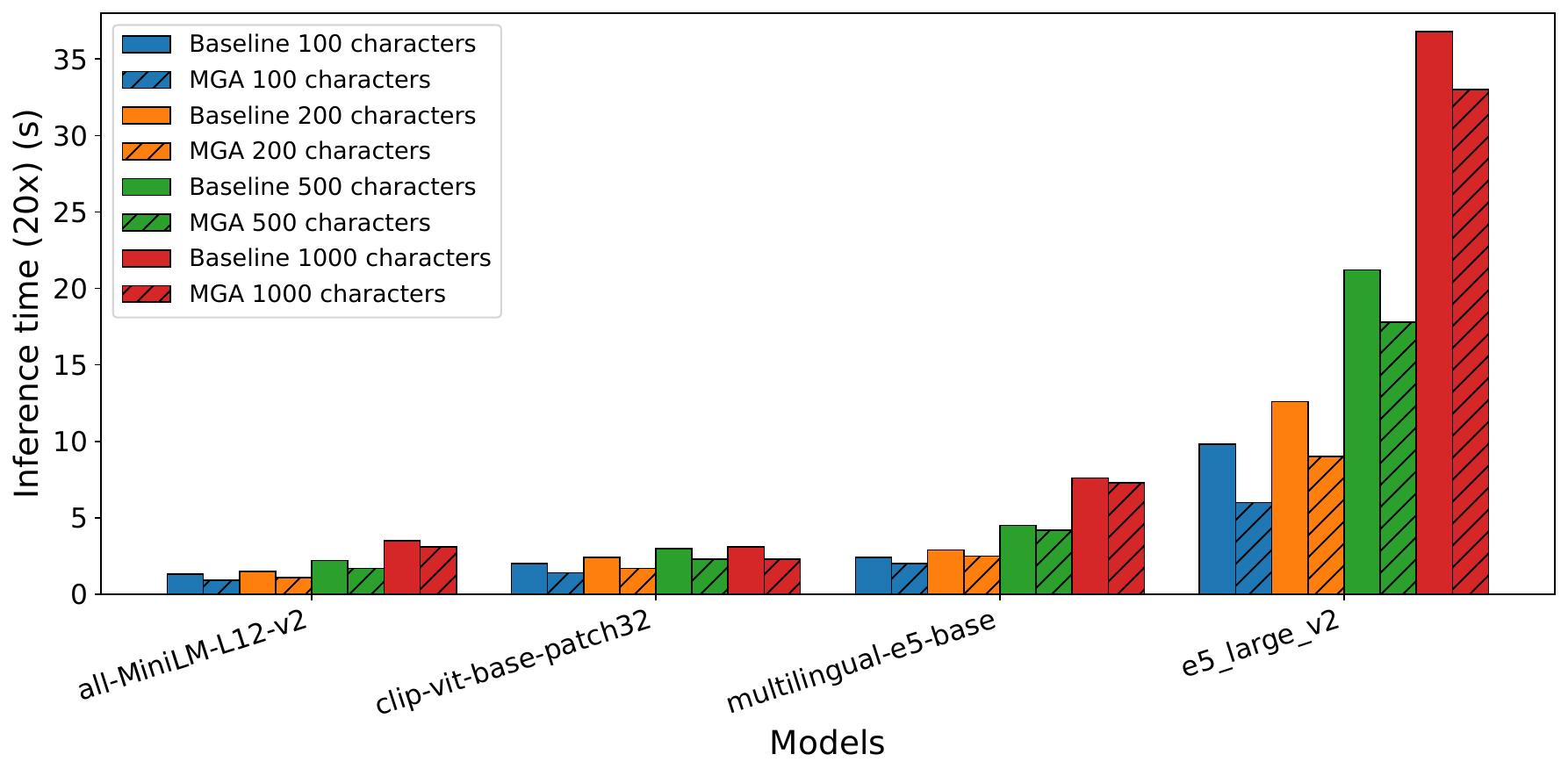}
  \caption{Runtime of ONNX model execution with and without weight sharing via MGA across input sizes.}
  \label{fig:onnx_inference}
\end{figure}

Figure \ref{fig:onnx_inference} reports the runtime behavior of ONNX inference as input length increases from 100 to 1000 characters, comparing the baseline configuration with the MGA sharing configuration. Across all models, inference latency increases monotonically with input length, reflecting the higher computational cost of processing longer sequences. However, ONNX model sharing through MGA consistently reduces runtime across all input sizes and models.

The magnitude of improvement varies with model size. For lightweight models such as \emph{all-MiniLM-L12-v2}, the absolute runtime reduction is modest, as inference cost is already low and less dominated by model initialization and setup overhead. In contrast, larger models incur substantially higher latency in the baseline configuration and thus benefit more from sharing. For \emph{e5\_large\_v2}, model sharing reduces inference latency by a large margin across all input lengths, with up to a 37\% reduction at shorter inputs, indicating that amortizing model loading and optimized weight preparation significantly lowers end-to-end execution time for heavyweight models.

Overall, Figure \ref{fig:onnx_inference} shows that ONNX model sharing through MGA is most effective when inference cost is non-trivial and model initialization overhead forms a significant fraction of execution time.

\subsection{Memory Footprint of ONNX Integration}

\begin{figure}
  \centering
  \includegraphics[width=0.8\linewidth]{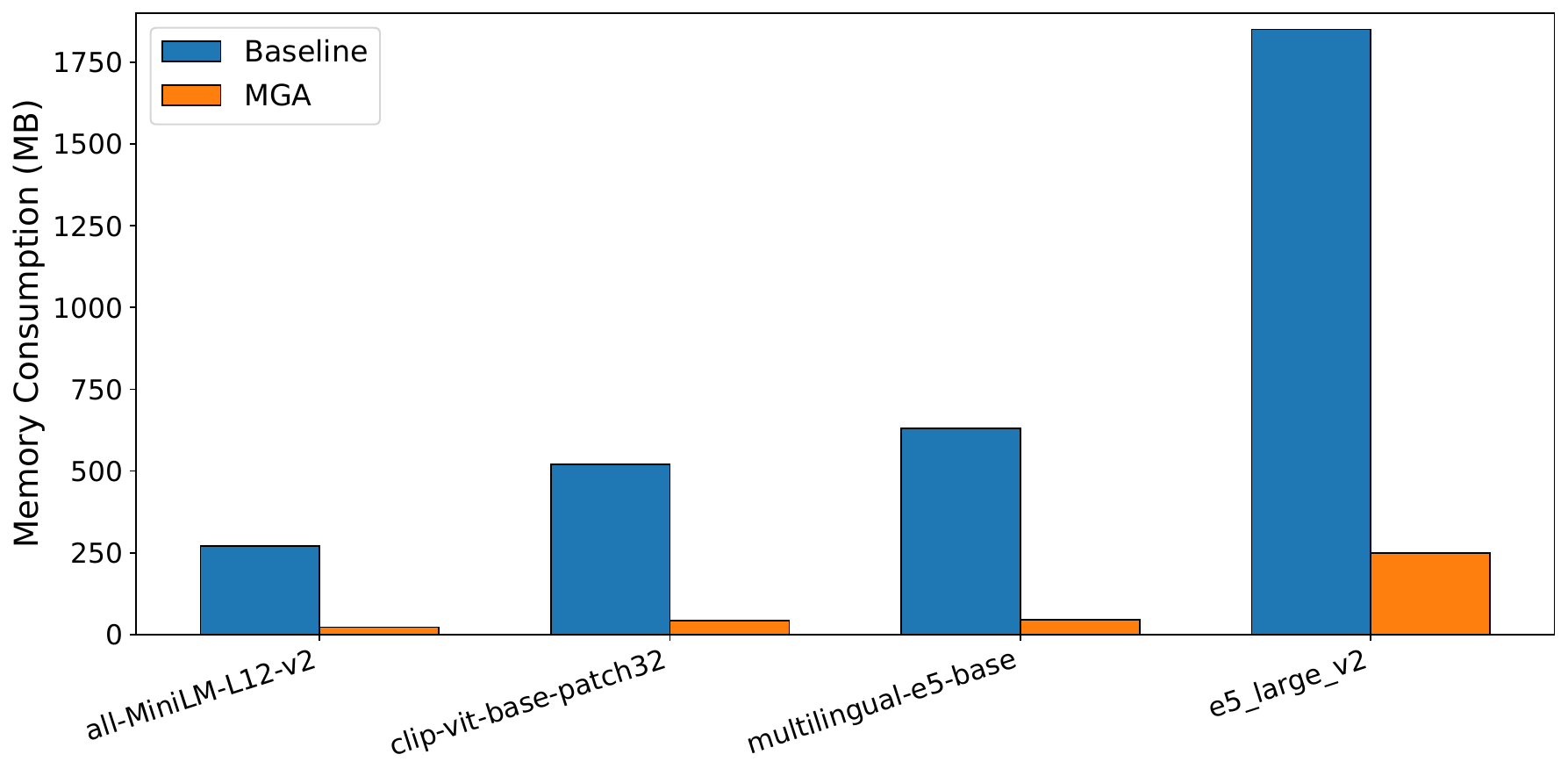}
  \caption{Memory footprint of ONNX model execution with and without model sharing via MGA.}
  \label{fig:onnx_memory}
\end{figure}

This section evaluates the memory footprint of ONNX model sharing using the same workload as Section \ref{runtime-overhead-of-onnx-model}, focusing on space efficiency rather than runtime behavior. Figure \ref{fig:onnx_memory} reports the peak memory usage under varying levels of concurrency. Without sharing, memory consumption is dominated by duplicated model graphs and optimized weight representations across concurrent processes. This results in high peak memory usage even for small models.

Across all evaluated models, sharing reduces peak memory usage by approximately
90\% to 96\%. The reduction is consistently large for both small and large models, indicating that memory savings are insensitive to model size. This is because model graphs and pre-optimized weights are materialized once and reused across concurrent executions, eliminating redundant per-process allocations.

These results show that, while runtime benefits depend on model size and inference cost, memory efficiency gains from ONNX model sharing are universal and predictable. This makes sharing particularly effective for high-concurrency in-database inference workloads and for large models that would otherwise impose significant memory pressure.

\section{Related Work} \label{related-work}
This section discusses work related to MGA on various Oracle memory management techniques and shared memory models from other database systems.
\subsection{Evolution of Memory Management in Oracle AI Database}
Memory management in Oracle AI Database has evolved significantly over time, reflecting a broader shift from static memory partitioning toward automated resource management. Traditional Oracle architectures divide memory into multiple regions based on their scope and lifetime, each serving distinct purposes \cite{dageville2002sql}. Early designs relied on manual configuration and static sizing of these regions, placing a substantial burden on database administrators.

To reduce manual intervention and better adapt to workload dynamics, Oracle introduced two memory management solutions:

\subsubsection{Automatic Memory Management (AMM)}
Oracle’s Automatic Memory Management treats SGA and PGA as a single budget that can expand or contract in response to load. A supervisory layer records how much of that budget belongs to the root container and how much has been delegated to each tenant, ensuring that the control plane can rebalance usage without breaking isolation. When pressure shifts from buffer caches to operator work areas, the manager shifts capacity accordingly yet the ownership ledger remains consistent.

\subsubsection{Automatic Shared Memory Management (ASMM)} ASMM in Oracle coordinates SGA resources by rebalancing SGA component sizes based on workloads. This is done via a centralized broker at the CDB level. 

The key difference between AMM and ASMM is that the latter is only available on hugetlbfs-backed SGA while AMM does not support large pages. ASMM is available for the SGA while AMM manages both SGA and PGA. Since MGA is accounted under PGA limits, it is automatically managed by AMM. Moreover, via MGA, the database can inherently manage its own memory better, for example, by releasing the backing memory to the OS dynamically.

\subsection{Shared Memory Management in Other DBMSs}

Other database systems adopt diverse strategies for shared and automatic memory management, reflecting different trade-offs in scope, control, and abstraction boundaries. 

PostgreSQL's Dynamic Shared Memory (DSM) \cite{postgresql_dsm_config_2023} supports parallel query execution by allowing worker processes to allocate and attach shared memory at runtime. While effective for query-level parallelism, DSM is query-scoped and relies on centralized metadata, limiting cross-query reuse and complicating failure handling. Other evaluated DBMSs do not have documented concepts of dynamically created, scoped shared memory.

SQL Server \cite{sqlserver_memory_model_2023} uses a centralized memory model with a shared buffer pool that dynamically allocates memory to subsystems based on workload. However, this design does not expose an abstraction for instance-scoped shared memory with explicit and configurable lifetime semantics.

In contrast to prior database systems, MGA provides an abstraction that explicitly separates memory scope and lifetime from specific execution contexts. This abstraction is implemented on customizable memory backends with memory allocation, sharing semantics, and lifecycle management (including instance state recovery) explicitly controlled within the database engine. Unlike existing approaches that restrict shared memory to a single query scope or rely on centralized synchronization structures, MGA exposes configurable scope boundaries that can extend across queries and system components.

\section{CONCLUSION AND FUTURE WORK} \label{conclusion}

We introduced the Managed Global Area (MGA), a scoped shared-memory abstraction in Oracle AI Database that enables dynamic, modular, configurable and process-selective memory sharing. MGA improves memory coordination across processes under strict production constraints. Evaluation on TPC-H workloads shows MGA reduces hash join latency by up to 35\%, with consistent gains across join-intensive queries. Among many other capabilities, it also supports in-database AI inference, reducing memory usage and lowering latency for large ONNX models in concurrent workloads.

MGA improves execution efficiency without compromising predictability or fault tolerance. It complements existing memory areas, bridging the gap between fully shared and private memory to match execution needs.

Future work will extend MGA's principles to broader environments, including cloud and heterogeneous memory systems.

\textbf{MGA Beyond Database Server}. One potential extension is to apply MGA to memory sharing between database server processes and local client connections. Oracle AI Database supports bequeath connections, in which a client and server instance running on the same host communicate using inter-process communication rather than a network listener. Leveraging MGA in this context could provide a shared memory substrate for exchanging query inputs, outputs, and execution state, reducing data movement overhead while preserving controlled access and isolation semantics.

\textbf{MGA with Trusted Execution Environment}. Another promising direction is the integration of MGA with trusted execution environments (TEE) to strengthen memory protection during execution. This is particularly relevant for Oracle external procedures, where an extproc agent runs as a separate process outside the core database engine. MGA could be used to efficiently share session or package state with extproc while avoiding repeated data copying. By combining MGA with capability-based access control within a trusted execution environment, access can be restricted to authorized memory regions, improving both security and performance when invoking external routines.

\textbf{Distributed shared memory}. Beyond single-instance deployments, MGA could serve as a foundation for distributed shared memory across multiple nodes. Prior work on distributed shared memory systems shows that exposing disaggregated memory through a shared abstraction can improve utilization and elasticity while tolerating data and workload skew \cite{wang_case_2022}. Extending MGA to operate over remote or disaggregated memory backends, such as RDMA-accessible or CXL-attached memory, could enable Oracle processes on different nodes to coordinate through shared regions with well-defined consistency and access semantics. Such an extension would preserve MGA’s flexibility while addressing the additional challenges of coherence, latency, and fault tolerance in distributed environments.

\begin{acks}
The authors gratefully acknowledge the teams and colleagues at Oracle for their valuable contributions to the collection of the performance results and data presented in this paper through their use of MGA.

\end{acks}

\balance
\bibliographystyle{ACM-Reference-Format}
\bibliography{sample}

\end{document}